\newcommand{\la}{\langle}
\newcommand{\ra}{\rangle}

\newcommand{\beq}{\begin{eqnarray}}
\newcommand{\eeq}{\end{eqnarray}}

\renewcommand{\theequation}{\thesection.\arabic{equation}}

\newcommand{\do}{^{\circ }}

\newcommand{\chis}{ChS\,\,}

\newcommand\con{\langle \bar {q} q \rangle }

\newcommand{\btem}{\bibitem}

\newcommand{\TK}{T.\ Kunihiro}

\newcommand{\MPTP}{Prog.\ Theor.\ Phys.}
\newcommand{\MPR}{Phys.\ Rev.}

\documentclass[seceq,supplement]{ptptex}
\usepackage[xdvi,dvips]{graphicx}
\usepackage{graphicx}
\usepackage{wrapft}
\begin{document}



\begin{center}
  Some Recent Topics on Possible Chiral Restoration
in Nuclear Medium
\\

Teiji \textsc{Kunihiro}%
\\
Yukawa Institute for Theoretical Physics, Kyoto University,\\
Sakyoku, Kyoto 606-8502, Japan
\end{center}



\abstract{
Some topics  are introduced on possible evidences of chiral restoration
in nuclei and related ones.
The topics include the 
$\sigma$-mesonic mode in nuclei, the vector mesons
in a nuclear matter, deeply bound pionic and
 Kaonic nuclei with a discussion on the 
nature of $\Lambda(1405)$.
Other related topics are briefly mentioned.
An emphasis is put on  that phenomena even in 
finite nuclei with normal and  sub-normal nuclear density,
as will be explored in the new project in GSI and possibly
in J-PARC,
are interesting for physics of finite-density QCD.
}


\setcounter{equation}{0}
\renewcommand{\theequation}{\arabic{section}.\arabic{equation}}

\section{Introduction}
According to the present 
understanding\cite{liq-gas,ktttt,hk94,BR,csc_rev},
the hadronic matter at relatively low temperature
might undergo various phase transitions successively
or simultaneously; the new phases and the transitions include
liquid-gas phase transition,
the nuclear $^1$S$_0$-, $^3$P$_2$- and 
$^3$S$_1$-$^3$D$_1$-superconductivity,
 the pion condensation, Kaon condensation,
H-dibaryon matter, a mixed state of hadron-quark 
phases, 
chiral restoration, color-super conducting (CSC)
 phases with various complications.
In the present report, focusing on the
chiral transition in nuclear matter,
I will discuss some characteristic changes in the scalar and vector
 correlations associated with
the (partial) restoration of chiral symmetry
 in the hadronic medium. Some recent topics 
will be also introduced,
which may have a relevance to the chiral properties
of finite nuclei\cite{chiral02}.
It will be clear that phenomena  in 
finite nuclei with normal and even sub-normal nuclear density,
as will be explored in the new project in GSI and possibly
in J-PARC,
are interesting as a physics of finite-density QCD.

\setcounter{equation}{0}
\renewcommand{\theequation}{\arabic{section}.\arabic{equation}}

\section{Chiral Properties at Finite Density}

Although there has been recently a tremendous development
in the lattice-QCD simulations at finite chemical potential
\cite{fodor},  it is  fair to say that such lattice simulations
are still premature to give 
a definite thing for the hadronic and/or quark matter,
 especially on a phase transition 
at finite density with relatively low temperatures.

A heuristic argument based on 
a Hellman-Feynman theorem can tell us that the chiral condensate 
$\con$ decreases at finite density $\rho_{_B}$ as well as at 
finite $T$.
For the the degenerate nucleon system $\vert {\rm NM}\ra$,
one may start from the formula\cite{DL90},
\beq
 \langle {\rm NM}\vert \bar{q}q\vert {\rm NM}\rangle=
\frac{\partial\langle {\rm NM}\vert {\cal H}_{QCD}\vert {\rm NM}\rangle}
{\partial m_q},
\eeq
where the expectation value of QCD Hamiltonian may be evaluated to 
be 
$ \langle {\rm NM}\vert {\cal H}_{QCD}\vert {\rm NM}\rangle=
\varepsilon_{vac}+\rho_{_B}[M_N+B(\rho_{_B})]$.
Here, $\varepsilon_{vac}, M_N$ and $B(\rho_{_B})$ denote the vacuum 
energy, the nucleon mass and the nuclear binding energy per particle,
respectively.
Thus one ends up with
\beq
{\langle {\rm NM}\vert\bar{q} q\vert {\rm NM} \rangle \over
 \langle \bar{q} q \rangle_{0} }
= 1 - {\rho_{_B} \over f_{\pi}^2 m_{\pi}^2 } \left( \Sigma_{\pi N}
 + \hat{m} {d \over d\hat{m}}  B(\rho_{_B}) 
\right) ,
\eeq
where
$\Sigma _{\pi N}=(m_u+m_d)/2\cdot\la N\vert \bar{u}u+\bar {d} d\vert 
N\ra$ denotes the  $\pi$-N sigma term with $\hat{m} =(m_u + m_d)/2$;
the semi-empirical value of $\Sigma _{\pi N}$ is known to 
be $(40 - 60)$ MeV\cite{piN_sig}.
Notice that the correction term with finite $\rho_{_B}$ is negative
and  gives a reduction of some
 30 - 50 \% of $\langle {\rm NM}\vert\bar{q} q\vert {\rm NM} \rangle $
already at the normal nuclear matter density $\rho_0 = 0.17 $fm$^{-3}$. 
One may notice that the physical origin of this reduction is clear;
the scalar probe given by the operator $\bar{q}_iq_i$
 hits either the vacuum or a particle  present in the system at
 $\rho_{_B}\not=0$, where
 $\bar{q}_iq_i$ picks up a positive contribution
to the chiral condensate  
because of the positive scalar charge $\la N | \bar{q}q | N \ra >0$
of a nucleon.

From the above estimate, one might suppose that
 the central region of heavy nuclei is dense enough to  cause 
a partial restoration of chiral symmetry, realizing
some characteristic phenomena of the chiral restoration in nuclear
medium, which may be observed by experiments 
in the laboratories on Earth\cite{chiral02}.

\setcounter{equation}{0}
\renewcommand{\theequation}{\arabic{section}.\arabic{equation}}

\section{The $\sigma$ mesonic mode in nuclei}

It is a well-known fact in many-body or statistical physics that
if a phase transition is of second order or weak first order,
there may exist specific  collective excitations called 
 {\em soft modes}\cite{soft}; they actually correspond to 
the quantum fluctuations of the order parameter.
In the case of chiral transition, 
there are two kinds of fluctuations; 
those of the phase and the modulus of the chiral condensate. 
The former is the Nambu-Goldstone boson, i.e., the pion, while 
the latter has the  quantum numbers $I=0$ and 
$J^{PC}=0^{++}$, which then may be identified with 
the meson historically called the $\sigma$ meson\cite{ptp85}.

\subsection{The scalar mesons in free space}

After  the establishment of the chiral perturbation theory
\cite{chipert} for
describing low energy hadron phenomena,
people has also come to 
be able to describe resonances
in a consistent way with chiral symmetry\cite{IAM,OOR}.
A central problem was recognized \cite{s-t-channel,crossing} to incorporate
the fundamental properties of the scattering amplitude
such as   unitarity,
 analyticity and the crossing symmetry together 
with chiral symmetry. 
Recent cautious phase shift
 analyses for the pi-pi scattering\cite{pipiyitp,pipi,crossing,CLOSETORN} and
 the decay processes of heavy particles such 
as D$\to \pi \pi \pi$
 \cite{E791}
 showed a  pole identified with
 the $\sigma$ in the $s$ channel together with the $\rho$ meson
pole in the $t$ channel:
The $\sigma$ pole has  the real part 
Re\, $m_{\sigma}= 500$-800 MeV and the imaginary 
part Im\, $m_{\sigma}\sim {\rm Re}\, m_{\sigma}$\cite{pipi,crossing}.

One should notice that there are serious 
controversies on the nature of the scalar mesons including 
the $\sigma$\cite{CLOSETORN}:
The low-lying scalar mesons with $J^{PC}=0^{++}$ of the simple
q-$\bar{\rm q}$ nature is at odd with
the conventional constituent quark model\cite{consti};
they should be a $P$-wave  state($^3P_0$), 
which is 
in turn usually heavier than 1.2 GeV.
Within the framework of the non-relativistic
constituent quark model,
the low-lying scalar mesons might be
described as diquark-anti-diquark states, i.e., four-quark states,
which can have as large as 600 MeV binding energy due to the 
color-magnetic interaction as argued by Jaffe\cite{jaffe}.
However, it should be noticed that the pion can not be understood
within the conventional constituent quark model, either.
As first shown by Nambu\cite{nambu},
 the pion may be interpreted as a {\em collective} state given as
a superposition of many $q$-$\bar{q}$ states.
Thus a natural interpretation of the $\sigma$
as the quantum
fluctuation of the chiral order parameter is also a collective
state as the pion as the phase fluctuation of the chiral 
order parameter \cite{nambu,ptp85}; the collectiveness of them
are  due to chiral symmetry and its dynamical breaking.
It may possibly be the case that the $\sigma$ pole 
is  only dynamically generated by the chiral dynamics, implying
that  the
$\sigma$ is a  $\pi$-$\pi$ molecule.
One may also mention that  the argument based on the 
``mended symmetry'' of Weinberg assumes the $\sigma$ mass 
below or equal to the $\rho$ meson mass.

\subsection{$\sigma$ meson in  hadronic matter}

If the $\sigma$ is really associated
with the fluctuation of the chiral order parameter,
the $\sigma$ can be
a {\em soft mode} of the chiral restoration
as was  first argued and demonstrated 
in \cite{ptp85}.
It implies that the $\sigma$ pole moves toward the origin of the
complex energy plane in the chiral limit and the $\sigma$ may become a sharp
resonance as chiral symmetry is restored 
 at high temperature and/or density;  see also
\cite{bernard87}.
The present author proposed  some experiments \cite{ptp95,tit}
to create the scalar mode in nuclei thereby to
obtain a clearer evidence of  the  existence of 
the $\sigma$ meson and also to examine the possible 
restoration of chiral  symmetry in the nuclear
 medium.

Notice, however, that a hadron ejected into a nuclear/hadron medium might
loose its identity and describing the whole system in terms of the 
(possibly) changed
 mass and width of the hadron can be inadequate. 
The most proper quantity to observe the behavior of a hadron 
in a matter is the response function or spectral function:
As long as  the coupling of the hadron  with the environment is relatively small,
a peak corresponding to the hadron 
remains with a small width in the spectral 
function, then one can speak of the width and the shifted mass
of the hadron in the matter.
The spectral function in the scalar channel is
obtained from the propagator. 
The $\sigma$-meson propagator at rest in the medium reads\cite{HKS}
$D^{-1}_{\sigma} (\omega)= \omega^2 - m_{\sigma}^2 - 
\Sigma_{\sigma}(\omega; \rho_{_B})$,
where $m_{\sigma}$ is the mass of the $\sigma$ in the tree-level, and
$\Sigma_{\sigma}(\omega; \rho_{_B})$ represents
the loop corrections
in the vacuum as well as in the medium.
 The corresponding spectral function is given by 
\beq
\rho_{\sigma}(\omega) = - \frac{1}{\pi} {\rm Im} D_{\sigma}(\omega).
\eeq
Now one can see that 
${\rm Im} \Sigma_{\sigma}\propto \theta(\omega - 2 m_{\pi})
 \sqrt{1 - 4m_{\pi}^2/ \omega^2}$
near the two-pion threshold  in the one-loop order.
On the other hand,
the pole mass $m_{\sigma}^*$ in the medium 
is  defined by
${\rm Re}D_{\sigma}^{-1}(\omega = m_{\sigma}^*)=0$.
Partial restoration of \chis implies that $m_{\sigma}^*$
  approaches to $ m_{\pi}$.  
Thus there should exist a density $\rho_c$ at which 
 ${\rm Re} D_{\sigma}^{-1}(\omega = 2m_{\pi})$
 vanishes even before the complete restoration
 of \chis where $\sigma$-$\pi$
 degeneracy gets realized;
\beq
{\rm Re} D_{\sigma}^{-1} (\omega = 2 m_{\pi}) =
 [\omega^2 - m_{ \sigma}^2 -
 {\rm Re} \Sigma_{\sigma} ]_{\omega = 2 m_{\pi}} = 0.
\eeq
At this point, the spectral function is solely 
given in terms of the
 imaginary part of the self-energy;
\beq
\rho_{\sigma} (\omega \simeq  2 m_{\pi}) 
 =  - {1 \over \pi \ {\rm Im}\Sigma_{\sigma} }
 \propto {\theta(\omega - 2 m_{\pi}) 
 \over \sqrt{1-{4m_{\pi}^2 \over \omega^2}}},
\eeq
which clearly shows the near-threshold enhancement of
the spectral function.
 This should be a general phenomenon to be realized
in association with  partial restoration of \chis.

To make the argument more quantitative,
Hatsuda, Shimizu and the present author \cite{HKS}
 evaluated the spectral function $\rho_{\sigma}(\omega)$ in 
 the O(4) linear $\sigma$-model and showed that the spectral
 enhancement near the $2m_{\pi}$ threshold takes place 
in association with  partial restoration of \chis  
at finite baryon density; the calculation was
 a simple extension of the 
finite $T$ case done by Chiku and Hatsuda\cite{CH98}.
 An improvement of this mean-field level calculation  was
subsequently made \cite{SCHUCK00} by including
 the $p$-$h$ and $\Delta$-$h$ contributions to the pion 
 propagator. In this case, the spectral strength
 spreads into the energy region even below 2$m_{\pi}$, but
 the qualitative feature of the enhancement at 2$m_{\pi}$
 still remains.

Interestingly enough, CHAOS collaboration  \cite{chaos} observed that 
the spectral function of the invariant mass  $M^A_{\pi^{+}\pi^{\pm}}$
for ${\pi^{+}\pi^{\pm}}$ from the reaction A($\pi^+, \pi^{\pm})$A'
  near the 2$m_{\pi}$ threshold with A ranging from $2$
to $208$\footnote{This experiment was motivated to explore the $\pi$-$\pi$
correlations in nuclear medium\cite{motivation}.}. 
They obtained the following interesting result:
When A$=2$, i.e., the target is deuteron, the spectral function
has only a tiny strength around the threshold, while it
 increases dramatically in the $I=J=0$ channel with increasing $A$.
The spectral function shows no such an increase in the $I=2$ channel.
For the experimental confirmation of 
 the threshold enhancement seen in \cite{chaos},
 measurements of  2$\pi^0$ and 
$2\gamma$ final states with hadron/photon beams off
 the  heavy nuclear targets are necessary.
 Those channels are free from the $\rho$  meson background
  inherent in the $\pi^+\pi^-$ measurement; see also \cite{CB00}.
An experiment detecting  2$\pi^0$ 
 from the reaction
 A$(\gamma, \pi^{0}\pi^{0})$A' with 
 $E_{\gamma}=400$ - 460 MeV and 
 A$=$ H, $^{12}$C, Pb has been made
by TAPS group\cite{taps}; as shown in Fig.1,
 a dramatical softening of the spectral
function in the $\sigma$ channel is seen 
for heavier nuclei. Notice that 
the pions are absorbed mostly in the surface region and 
may be difficult to prove the interior of nuclei,
while the $\gamma$ is 
more easily penetrate deep into  nuclei and can prove the
possible density effects on the spectral function in the
matter as shown 
in \cite{roca}\footnote{However, see a recent work \cite{mosel} 
in which  an account is given of the downward shift 
of the $\pi^0\pi^0$ spectral function 
in terms of the conventional final state interactions of the pions 
with the nucleons. Clearly, more theoretical and experimental works
are needed to settle down the issue.}
\begin{figure}[htbp]
\centering
\includegraphics[scale=.55]{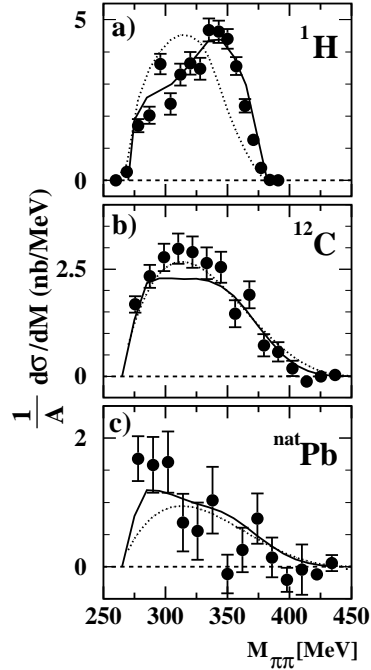}
\caption{
Differential cross sections of the reaction
 A$(\gamma, \pi^{0}\pi^{0})$A' 
with  A$=$ H, $^{12}$C, Pb for incident 
photons in the  energy range of 400 - 460 MeV.
Taken from \cite{taps}.
}\label{fig:taps}
\end{figure}
One can also study the in-medium
$\pi$-$\pi$ amplitude   using the same O(4) model:
Jido, Hastuda and the present author \cite{jhk} 
showed that a large enhancement of the cross section 
 near the threshold is obtained  along with the increase of the baryon density.
Then one may ask 
 whether the near-threshold enhancement obtained in the O(4)
 {\em linear} $\sigma$ model can 
arises also in the {\em non-linear} models which lack in
the explicit $\sigma$ field. 
In  \cite{jhk}, it was shown that  it is the case and
also an vertex in the non-linear chiral Lagrangian is identified 
which is responsible for the enhancement.

 
 To study these problems, it is best to start with
 the standard polar parameterization of the chiral field\cite{jhk},
 $M = \sigma + i \vec{\tau} \cdot \vec{\pi}
  = (\la \sigma \ra + S) U $ with $U = \exp (i \vec{\tau}
 \cdot \vec{\phi} /f^{*}_{\pi})$. Here,
$\la \sigma \ra$ is the chiral  condensate in nuclear matter
 as before, while 
 $f^{*}_{\pi}$ is to be the  ``in-medium pion decay constant''.
 Taking the heavy-scalar and heavy-baryon limit 
limit simultaneously, and integrating out 
 the scalar field  $S$, one can 
obtain the following effective Lagrangian:
\beq
\label{model-nl2}
{\cal L}_{\rm eff}  = 
 \left(
 {f_{\pi}^2 \over 4}
 - {g f_{\pi}  \over 2 m_{\sigma}^2}\bar{N}N
 \right)
\left(
{\rm Tr} [\partial U \partial U^{\dagger}]
 - {h \over f_{\pi}} \  {\rm Tr}[U^{\dagger}+U] \right)
 \ + \  {\cal L}_{\pi N}^{(1)} + \cdot \cdot \cdot \ \ ,
\eeq
where
In (\ref{model-nl2}), 
${\cal L}_{\pi N}^{(1)}$ is the standard $p$-wave $\pi$-$N$ coupling
 and  $\cdot \cdot \cdot$  denotes
  other higher dimensional operators which are not relevant for the
   present discussion.
Here all the constants are taken for the vacuum:
 $f_{\pi} = \la \sigma \ra_0 $,
 $m_{\sigma}^2 = \lambda \la \sigma \ra_0^2/3 + m_{\pi}^2$,
 and  $m_N = g \la \sigma \ra_0$.

\begin{figure}[tb]
\begin{minipage}{.5\linewidth}
\centering
\includegraphics[scale=.5]{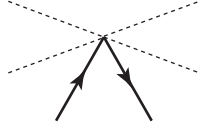}
\end{minipage}
\begin{minipage}{.35\linewidth}
\caption{
The new 4$\pi$-$N$-$N$ vertex generated
 in the nonlinear realization.
 The solid line with arrow and the dashed line represent
 the nucleon and pion, respectively.}
\label{fig-vertex}
\end{minipage}
\end{figure}

$\bar{N}N$ in  eq.(\ref{model-nl2}) may be replaced by the baryon density 
$\rho$ in the mean-field approximation, leading to 
 a reduction of  the vacuum condensate;
\beq
f_{\pi} = \la \sigma \ra_0  \rightarrow
  \la \sigma \ra =
 \la \sigma \ra_0 (1- g  \rho/f_{\pi} m_{\sigma}^2)
 = f_{\pi}^*.
\eeq
This implies that  the proper normalization of the
pion field  in the  nuclear medium should be
\beq
\phi' = (\phi /f_{\pi}) \cdot f_{\pi}^* \equiv {Z^{\ast}}^{1/2}\phi,
\eeq
with ${Z^{\ast}}^{1/2}\equiv f_{\pi}^*/f_{\pi}$.
This is the wave-function renormalization in the medium. 
One can now see
that the origin of the  near-threshold 
enhancement is the wave-function renormalization
in the nuclear medium,  ascribed to the following  new vertex;
\beq
\label{new-vertex}
{\cal L}_{\rm new} = - {3g \over 2 \lambda f_{\pi}} \
\bar{N}N {\rm Tr} [\partial U \partial U^{\dagger}].
\eeq
owing to the scalar nature,
this vertex can affect not only the pion propagator  but also
 the interaction among pions in the nuclear medium.
 In Fig.\ref{fig-vertex},  
4$\pi$-$N$-$N$ vertex generated by ${\cal L}_{\rm new}$  is shown.

Here it should be pointed out that
 the vertex eq.(\ref{new-vertex})
 has been known to be one of the next-to-leading order terms
 in the non-linear chiral Lagrangian in the
heavy-baryon formalism  \cite{GSS88}; see \cite{mow} for a recent
 development
on the in-medium chiral perturbation theory.


\subsection{Simultaneous softening of the $\sigma$ and 
the $\rho$ mesons associated with chiral restoration}

Recently, Yokokawa et al\cite{yokokawa} have constructed a
unitarized $\pi$-$\pi$ scattering amplitude in 
a  hot and dense matter, and 
 shown that the spectral functions in the $\sigma $ 
and the $\rho$ meson channels gives rise to a softening
 {\em in tandem} as the chiral
 symmetry is restored.
They adopted the $N/D$ method\cite{ND} a la Igi and
Hikasa\cite{crossing}, in which 
 the scattering amplitude satisfies the analyticity and approximately
the crossing symmetry.
 The effect of the effect of chiral restoration
is taken into account in the mean field level, which 
is  tantamount to replacing $f_{\pi}$ by $f_{\pi}^*(\rho)$.

Four types of chiral models were taken to construct the scattering amplitude which 
is to be unitarized to check the possible model dependence:
Model A; The ``$\rho$ model'' in which the $\pi$ and bare-$\rho$ are the basic
fields, and the $\sigma$ is to be generated dynamically.
\, Model B;The ``$\sigma$ model'' in which the $\pi$ and bare-$\sigma$ are the basic
fields, and the $\rho$ must be be generated dynamically.
\, Model C;The  ``degenerate $\sigma$-$\rho$ model'' in 
which the $\pi$, bare-$\sigma$ and bare-$\rho$ are all the basic
fields.
\, Model D; The leading chiral Lagrangian ${\cal L}_2$, which solely
can generate the $\sigma$ dynamically but not the physical $\rho$.

In Fig.3,
the trajectories of the moving complex poles in $I=J=0$ channel along with 
the decrease of $f_{\pi}^{\ast}$ are
shown for the four models; the crosses indicate the pole positions in the
vacuum. One sees that the  
 It should be remarked that there exist two poles in the both channels (except
for Model D), one of which moves toward the
$2m_{\pi}$ threshold. 
It implies that the sigma meson which is elusive in the free space may
appear as a rather sharp resonance in hot and/or dense medium
 where chiral symmetry is partially restored. This pole behavior was
first suggested in \cite{jhk} and shown in \cite{paris}.
A remarkable point is that the softening also occurs 
in  the $\rho$ meson channel, which leads to some interesting implications
to the in-medium cross sections, which are shown in Fig. 4.
\begin{figure}[htbp]
\begin{minipage}{.5\linewidth}
\includegraphics[scale=0.4]{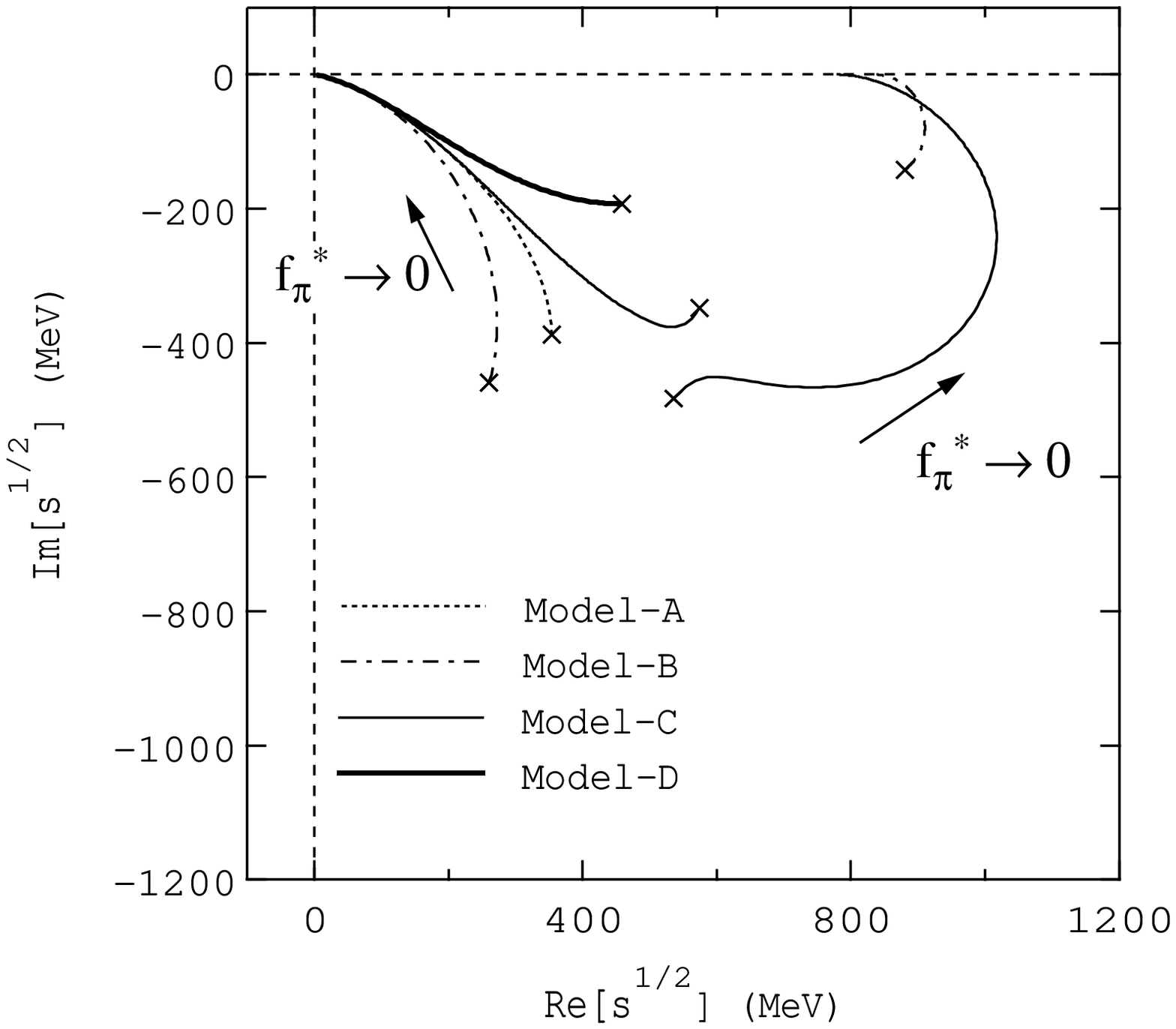}
\end{minipage}
\begin{minipage}{.5\linewidth}
\includegraphics[scale=0.4]{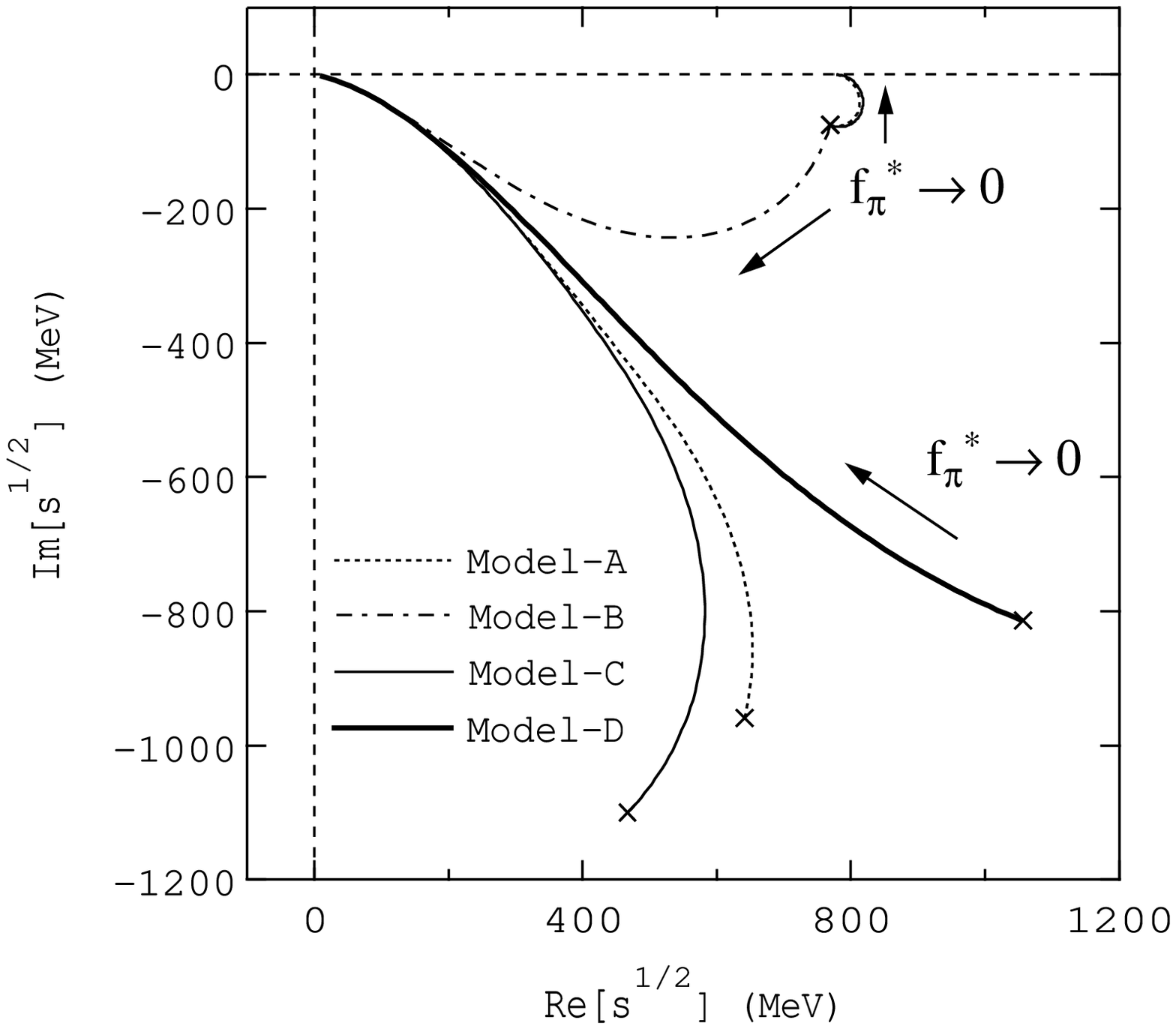}
\end{minipage}
\caption{\label{fig:pole} The movement of the poles in 
$I$=$J$=0  channel (left panel) and $I$=$J$=1  channel (right panel)
along with the decrease of $f_{\pi}^*$.
The crosses  are the pole positions in the vacuum.
Taken from \cite{yokokawa}.}
\end{figure}
The upper (lower) panels are for the case with small (large)
restoration of chiral symmetry
 with  $0.5 f_\pi < f_\pi^* < f_\pi$ ($0.1 f_\pi < f_\pi^* < 0.5 f_\pi$),
which is supposed to correspond to lower (high) densities.
At low densities, one sees that a significant of the red-shift of the peak
 (softening) of the cross section occurs in the $\sigma$ meson channel,
while a large broadening with a relatively small softening occurs in 
the $\rho$ meson channel, which, as will  be discussed in the subsequent section,
 has a strong relevance to the spectral
change obtained from the relativistic heavy ion collisions by
 CERES collaboration\cite{CERES01}.

\begin{figure} [htbp]
\begin{minipage}{.5\linewidth}
\includegraphics[scale=0.4]{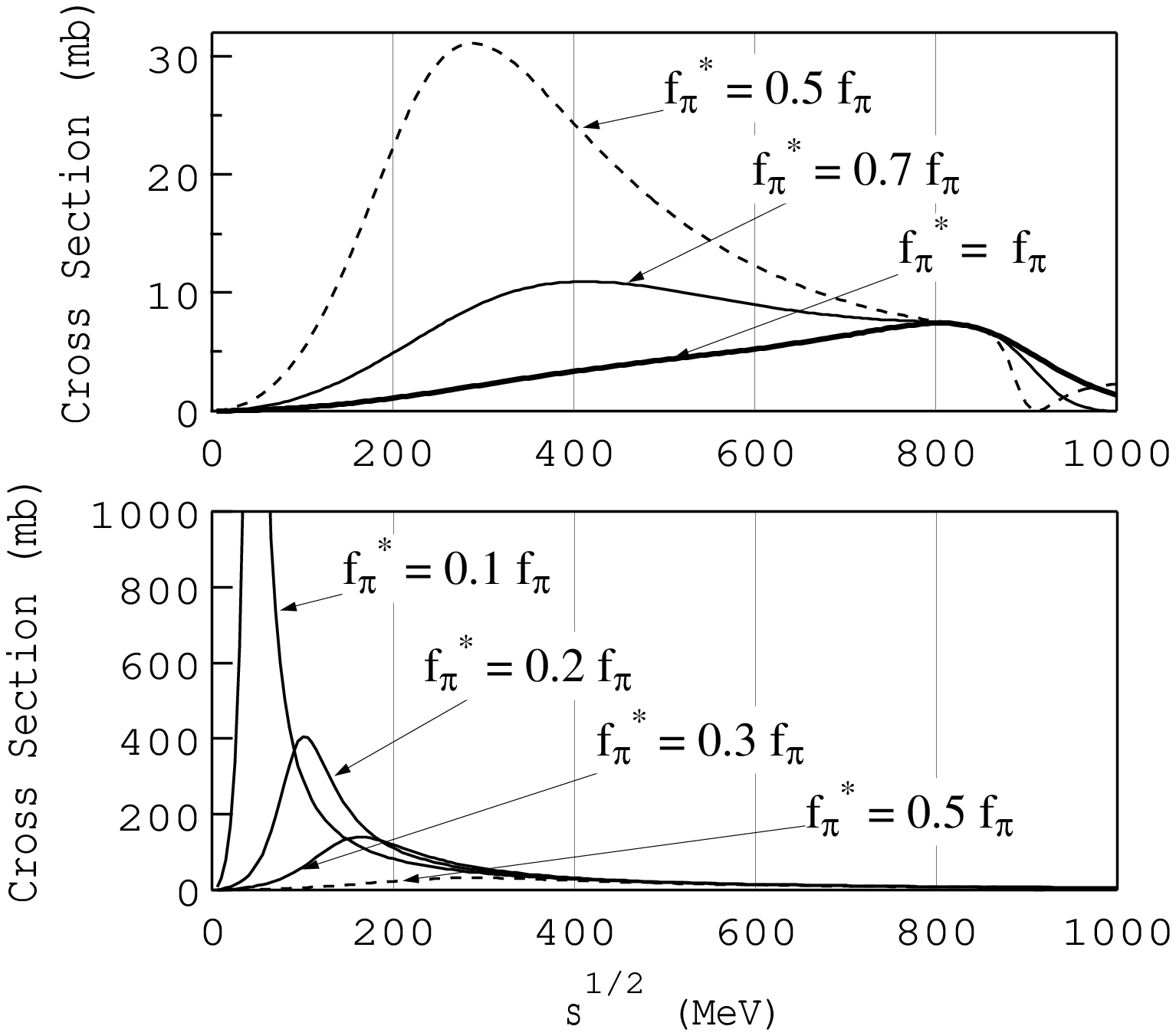}
\end{minipage}
\begin{minipage}{.5\linewidth}
\includegraphics[scale=0.4]{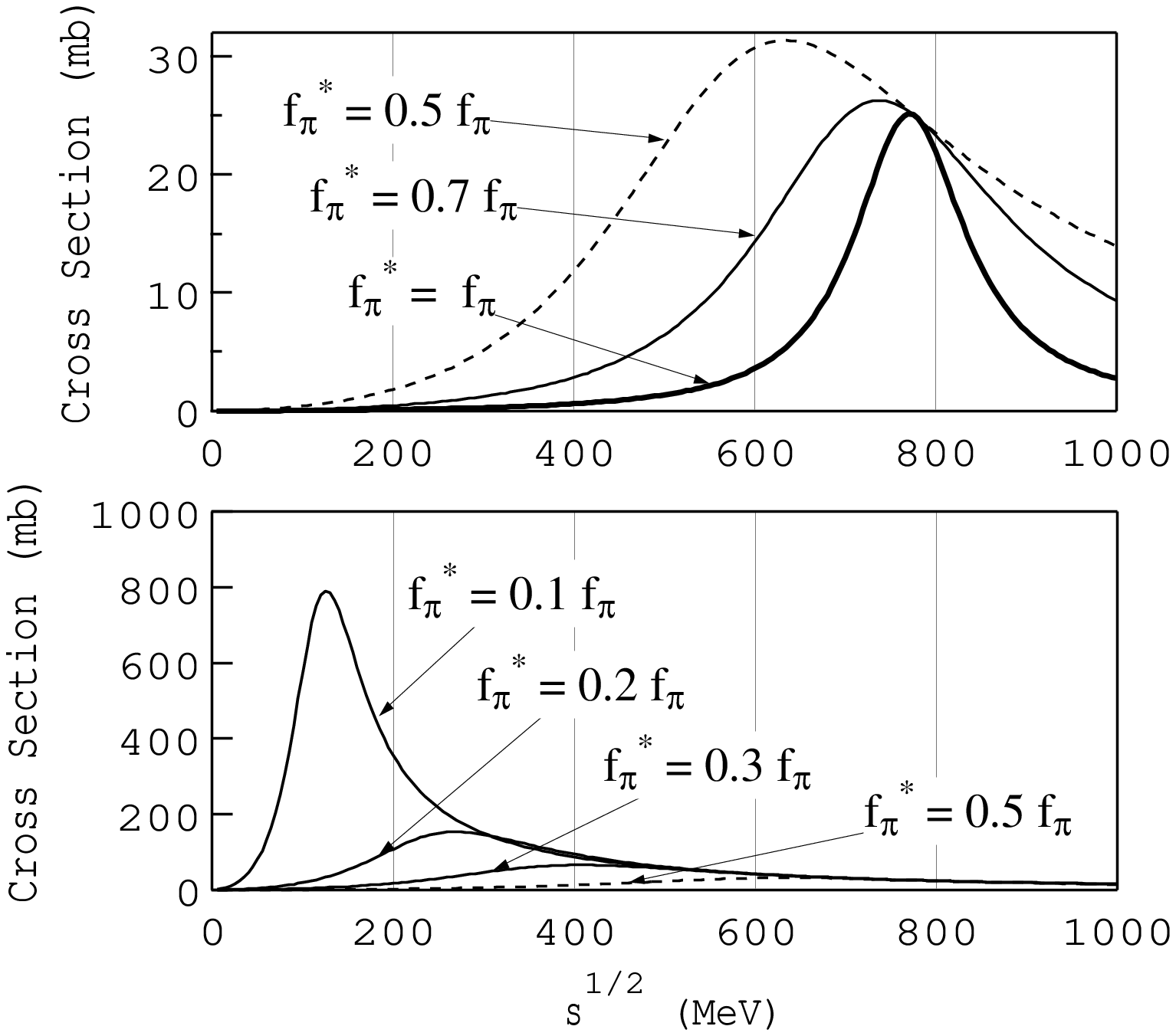}
\end{minipage}
\caption{\label{fig:test}
       The in-medium $\pi$-$\pi$ cross sections in $I$=$J$=0 channel
(left panel) and $I$=$J$=1 channel (right panel)  for  Model-B:
 Each upper (lower) panel shows the case of small (large) restoration
 corresponding to 
$0.5 f_\pi < f_\pi^* < f_\pi$ ($0.1 f_\pi < f_\pi^* < 0.5 f_\pi$).
Taken from \cite{yokokawa}.}
\end{figure}

\section{Chiral restoration and vector mesonic modes  in nuclei}

The spectral function deduced from the lepton pairs from the
heavy ion collisions such as Pb-Au collisions
 both at high energy (158 GeV/A) and at low energy (40 GeV/A)
 show a sizable enhancement of the $e^+e^-$ yield
 below the $\rho$-meson peak  \cite{CERES01}.
 This may or may not be 
 related to the partial chiral restoration in nuclear
medium originally proposed in \cite{P-BR-HL}.
It is worth mentioning that a simultaneous softening of the spectral function in 
the $\sigma$ meson (or $f_0(600)$) and the
$\rho$ meson discussed in the last section\cite{yokokawa}
 may also account for the spectral change seen in 
the STAR experiment at RHIC\cite{star}, as argued in
\cite{shuryak03}. In the STAR experiment\cite{star},
observed signals are reported
 for a variety of mesonic as well as baryonic resonances  produced by 
mid-central Au-Au collisions at $\sqrt{s}=200$ GeV, where
 the temperature effect should dominate the density effect.

 The E325 experiment at KEK  \cite{OZAWA01}  measured
 $e^+e^-$ pairs from the p-A collision at 12 GeV.
 The similar enhancement over the known source and
 combinatorial background as CERES is seen in the mass range
 of about 200 MeV below
 the $\rho$-$\omega$ peak for A$=$Cu.

\section{Deeply bound pionic atoms and reduction of $f_{\pi}^{\ast}$ in nuclei}

It is interesting that there are other possible 
experimental evidences for partial chiral restoration in 
nuclear matter than the chiral fluctuations in the
sigma meson channel discussed so far.
  The deeply bound pionic atom has proved 
to be a good probe of the properties of the hadronic interaction
 deep inside of heavy nuclei.  It has been 
 suggested \cite{WEISE,itahashi} that
the anomalous energy shift of the pionic atoms (pionic nuclei)
owing to the strong interaction could be attributed to the
 decrease of the effective pion decay constant 
$f^{\ast}_{\pi}(\rho)$ at finite density $\rho$ which 
may imply that the chiral symmetry is partially restored deep
inside of nuclei. This scenario has been confirmed \cite{kkw} by a
microscopic calculation taking into account the
energy dependence of the pion optical potential. It is interesting
that  Kolomeitsev et al\cite{kkw}
 also showed that their  result can be  understood 
in terms of the in-medium renormalization of the wave 
function\footnote{However, see also 
a recent phenomenological analysis \cite{gal04}, where
it is argued that  the energy dependence of the on-shell pions
is more essential, as in \cite{teoericson}.},
 as was important in the softening of the spectral function in 
the $\sigma$ channel\cite{jhk}.

\section{The nature of $\Lambda(1405)$ and possible deeply bound Kaonic nuclei}
 
The anti-Kaon in nuclear matter show a very attractive
nature. Waas and  Weise \cite{waas} showed that the spectral function for
the anti-Kaon in nuclear matter shows a dramatical softening, as
 shown in Fig.1 and 2 in their paper\cite{waas}.
Akaishi and Yamazaki\cite{aka-yama} considered seriously 
the attractive nature of the Kaon-nuclear
interaction which may lead to the  $\Lambda(1405)$ as  a
bound state, and showed that  the 
attraction can manifest itself in  a more dramatic way  
in a proton-rich matter.
They demonstrated that the Kaonic nuclei could be deeply bound
matter.
The creation of such an abnormal  state of matter has been also
confirmed by Dote et al\cite{dote} using an ab initio molecular dynamical
calculation called Anti-symmetrized molecular dynamics 
developed by Horiuchi\cite{amd}.
They showed that the created matter may have as high as 8 times
of the normal nuclear density.
After the workshop, a report was made of the possible evidence
of such an exotic state\cite{aka-yama-iwa}.
If such a dense matter with strangeness 
was confirmed, it would open a new era of nuclear/hadron physics.
It provides systems with high density and strangeness.
It can be a laboratory on Earth for examining the dense
matter expected in the interior of 
neutron stars and possibly in quark stars.

\section{Summary and concluding remarks}

I have tried to convince you that the nuclear matter at and even below
the normal nuclear density ($\rho\simeq \rho_0$) would be also
interesting  for physics of finite density QCD:
\begin{enumerate}
\item One might observe  precursory phenomena of chiral restoration
in various channels, i.e., 
 the $\sigma$- and the vector meson  channels and also the baryon
channels\footnote{I failed to give an account of it owing to the lack
of time.} 
\item
Such a research with the matter at $\rho\simeq \rho_0$ can give
a sound basis for attacking and exploring high-density matter,
for example, through the study on chiral dynamics at $\rho\not=0$.
\item 
Most importantly, rich experimental information will be available
in the near future, from the ongoing and future projects in GSI,
 KEK and J-PARC (former JHF).
\end{enumerate}

There are several topics which are related to chiral restoration in the
nuclear medium but  were not covered in the
present report because of the lack of time.
These include: \\
(i)\, The chiral properties of baryons of positive and negative
parity and their behavior along with the chiral restoration\cite{baryon}. 
(ii)\, The effects of  the scalar and vector
 correlations at $\mu_B\not= 0$, which may 
manifest itself in a peculiar behavior in 
the scalar and vector susceptibilities at 
$\mu\not=0$, especially 
 around the critical end point of the chiral transition
\cite{kuni91}.
(iii)\, Possible influence of 
the possible precedent  meson condensation and H-baryon matter to
the QCD phase transition at finite density\cite{tama}.

In passing, I would like to  emphasize that the 
experimental data from the facilities and theoretical works
based on effective theories and lattice QCD will cultivate the
field of finite density QCD together with astronomical data from
neutron stars, supernovas and hopefully quark stars.

\section*{Acknowledgements}
Major part of the present report 
is based on the works done in collaboration with T. Hatsuda, 
K. Hayashigaki,  D. Jido,  H. Shimizu and
K. Yokokawa, to whom the author is grateful.
This work is supported by the Grants-in-Aids of the Japanese
Ministry of Education, Science and Culture (No. 14540263). 
%

\end{document}